
\magnification 1200
\baselineskip 22pt
\noindent
To be published in Ann. of NYAS,{\it Fundamental Problems in
Quantum Theory}, D. Greenberger, (ed.)  1994;
hep-th/9411196

\vskip .8cm
\centerline{\bf  PROTECTIVE MEASUREMENTS}
\vskip .8cm
\centerline{ \bf Yakir Aharonov$^{a,b}$ and Lev Vaidman$^a$}

\vskip 1cm

\centerline{\it $^a$  School of Physics and Astronomy}
\centerline{\it Raymond and Beverly Sackler Faculty of Exact Sciences}
\centerline{\it Tel-Aviv University, Tel-Aviv, 69978 ISRAEL}

\vskip .5cm

\centerline{\it $^b$ Physics Department, University of South Carolina}
\centerline{\it Columbia, South Carolina 29208, U.S.A.}

\vskip 1.5cm

Protective measurements yield properties of the quantum state of a single
quantum system  without affecting  the quantum state.
A protective measurement involves adiabatic  coupling to the
measuring device together with a procedure to protect the state from
changing.
For nondegenerate energy eigenstates the protection is provided by the
system itself. In this  case it is actually possible to measure the
Schr\"odinger
wave via   measurements  on a single system. This fact provides an
argument in favor of associating physical reality with a quantum state of a
single system, challenging the usual ensemble interpretation. We also believe
that the complete description of a quantum system
requires a two-state vector formalism involving  (in addition to the usual one)
a future quantum state evolving backwards in time. Protective measurements
testing the two-state vector reality are constructed.

 \vskip 2cm

\break

\centerline{\bf INTRODUCTION}

Recently, we have proposed {\it protective measurements}$^{1,2}$ that allow
measuring the Schr\"odinger wave of a single particle.  We have argued that the
possibility
of such measurements tells us that a quantum state has more  physical
meaning than  is usually assumed; that is,  the Schr\"odinger wave is real in
some sense. A quantum state is not only a
statistical property of an ensemble, it is a property of
a single system.

  Also, in recent years we
developed an approach in which a quantum system is described, at a given
time,  by two (instead of one) quantum states: the usual one
evolving toward the future and the second evolving backwards in time from a
future measurement.$^{3-8}$ This approach proved itself fruitful at least for
describing measurements performed on pre- and post-selected ensembles.
In this approach, the vector describing a quantum system at a given time
consists
of two states.

 The following questions  arise: Is there a contradiction between these two
approaches? Which description is  appropriate: the standard, single-state, or
our two-state description?  Does the two-state vector have physical meaning
for a single system?  Is it  possible to measure this vector on
a single system? In this work we shall try to answer these questions.

In the following section  we present  our method of protective measurements of
a
single quantum state. This is  followed by a  brief review of the two-state
vector formalism. Then,  the main result of this work --  the method of
protective measurement
of a two-state vector -- is presented. We  conclude with a  discussion of the
obtained results.
\vfill

\break
\vskip .7 cm

\centerline{\bf MEASUREMENT OF THE SCHR\"ODINGER WAVE OF A SINGLE PARTICLE}

\vskip .2 cm

 At present, the commonly accepted interpretation of the Schr\"odinger wave is
due to Born.  He proposed to interpret the wave intensity not as the
density of distribution of actual matter, as Schr\"odinger first imagined, but
as
a probability density for the presence of a particle. Schr\"odinger,
however, wanted to believe that his wave represents a single particle: the
wave is an extended object really moving in space.   Born's
interpretation was supported by the fact that nobody knew how to measure the
density of the Schr\"odinger wave on a single system. There was a general
belief that the Schr\"odinger wave could only be tested for an ensemble of
particles. We have proposed  a new type of measurements: ``protective
measurements'' that
allow direct measurement of the  Schr\"odinger wave density on a
single particle. We have shown that one  can
simultaneously measure the density and the current of the Schr\"odinger
wave in many locations. The results of these measurements then allow the
reconstruction of the Schr\"odinger wave.

The simplest protection procedure is introducing a protective potential  such
that the
quantum state of the system will be a nondegenerate eigenstate of the
Hamiltonian. In fact, in many important cases this protection is given by
nature: almost isolated systems  will eventually decay to their  ground state
or to  some stable excited
state.

As an example of a simple protective measurement, let us consider a particle in
a discrete nondegenerate energy
eigenstate $\Psi (x)$. The standard von Neumann
procedure for measuring the value of an observable $A$ in
this state involves an
 interaction Hamiltonian,
$$
 H = g(t) P A,\eqno(1)
$$
coupling the system to a measuring device, or pointer, with coordinate
and momentum denoted, respectively, by $Q$ and $P$.  The time-dependent
coupling $g(t)$ is normalized to $\int g(t) dt =1$, and  the initial state
of the pointer is taken to be a Gaussian centered around zero.

In standard impulsive measurements, $g(t) \neq 0 $  for only a very
short time interval.  Thus, the interaction term dominates the rest of
the Hamiltonian, and the time evolution
$
e^{-{i\over \hbar}P A}
$
leads to a correlated state: eigenstates of $A$ with eigenvalues $a_n$
are correlated to measuring device states in which the pointer is
shifted by these values $a_n$.  By contrast, the protective
measurements of interest here utilize the opposite limit of extremely
slow measurement.  We take $g(t) = 1/T$ for most of the time $T$ and
assume that $g(t)$ goes to zero gradually before and after the period
$T$. We choose the initial state of the measuring device  such
that the canonical conjugate $P$ (of the pointer variable $Q$) is
bounded.  We also assume that $P$ is a constant of motion not only of the
interaction Hamiltonian (1), but of the whole Hamiltonian. For $g(t)$ smooth
enough we obtain an adiabatic process in
which the particle cannot make a transition from one energy eigenstate
to another, and, in the limit $T \rightarrow \infty$, the interaction
Hamiltonian does not change the energy eigenstate.  For any given value of
$P$, the energy of the eigenstate shifts by an infinitesimal amount
given by first order perturbation theory:
$$\delta E = \langle H_{int}  \rangle  = {{
\langle
A\rangle P}\over T}.
\eqno(2)$$The corresponding time evolution $ e^{-i P \langle A\rangle /\hbar} $
shifts the
pointer by the average value $\langle A
\rangle $.  This  result contrasts with the usual (strong)
measurement in which the pointer shifts by one of the eigenvalues of
$A$.
  By measuring the averages of a sufficiently large number of
variables $A_n$, the full Schr\"odinger wave $\Psi (x)$ can be
reconstructed to any desired precision.

As a specific example we take the $A_n$ to be (normalized) projection
operators on small regions $V_n$  having volume $v_n$:

$$A_n=\cases{ {1\over {v_n}},&if $x \in V_n$,\cr
0,&if $x \not\in V_n$.\cr} \eqno(3)
$$

\noindent
 The measurement of $A_n$ yields $$\langle A_n \rangle = {1\over {v_n}}
\int_{V_n} |\Psi|^2 dv = |\Psi_n|^2 ,\eqno(4) $$ where $ |\Psi_n|^2 $
is the average of the density $\rho(x) = |\Psi(x)|^2$ over the small
region $V_n$.  Performing measurements in sufficiently many regions
$V_n$ we can reconstruct $\rho(x)$ everywhere in space.  (Simultaneous
measurement of all the variables $A_n$ requires slower and weaker
interactions, and thus takes more time.)  For a real state the density
$\rho(x)$ is itself enough to reconstruct the Schr\"odinger wave; we
can fix the sign by flipping it across nodal surfaces.

In the general case, however,  we have to measure current
density in addition
to measurements of the density $\rho(x)$.  This time we also adiabatically
measure the averages of
$$
B_n ={1\over{2i}} (A_n\nabla + \nabla A_n)  ~~~.\eqno(5)
$$
Indeed, $\langle B_n\rangle$ are the average values of the current
$j = {1\over{2i}} (\Psi^* \nabla \Psi - \Psi  \nabla \Psi^* )$ in the
region $V_n$. Writing $\Psi (x) = r(x) e^{i\theta(x)}$ with
$r(x)=\sqrt{\rho (x)}$,
 we find that
$$
{{ j(x)} \over {\rho (x)}} = \nabla \theta ~~~;\eqno(6)
$$
and  the phase $\theta (x)$ can be found by integrating
$j/\rho$.

For a charged particle, the density $\rho (x)$ times the charge yields
the effective charge density. In particular, it means that an
appropriate adiabatic measurement of the Gauss
flux out of a certain region must yield the expectation value of the
charge inside this region (the integral of the charge density over
this region). Likewise, adiabatic measurement of the Ampere contour integral
yields the expectation value of the total current flowing through this
contour in the stationary case.

Our procedure is not applicable to  degenerate energy eigenstates.
The simplest way to deal with this case is by adding a potential (as
part of the measuring procedure) to lift the degeneracy. This
protection does not change the state. However, one can argue that it
changes the physical situation. We can bring this change to a minimum
by
 adding strong protection potential for a dense set of very short time
intervals. Thus, most of the time the system has not only the
same state, but also the original potential.

 We can measure even a superposition of energy
eigenstates by a similar procedure.  We
 add a dense set of time-dependent potentials acting for very
short periods of time such that the state at all these
times is the eigenstate of the Hamiltonian together with the
additional potential. Still, most of the time the system evolves under
the free Hamiltonian. The proof of the efficiency of the above strong
impulsive potentials is similar to the proof of the Zeno ``paradox"
in which a quantum system under a dense set of observations
 evolves
in accordance with the evolution tested, and not according to the
free Hamiltonian. In our case, the two evolutions are identical.
\vskip .7 cm

\centerline{\bf TWO-STATE VECTOR DESCRIPTION OF A QUANTUM SYSTEM}

\vskip .2 cm

In 1964 Aharonov, Bergmann and Lebowitz$^9$ considered measurements performed
on
a quantum system between two other measurements, results of which were
given. They proposed  describing the quantum system  between two
measurements by using two states: the usual one, evolving towards the future
from the time of the first measurement,
and a second state  evolving backwards in time, from the time of the second
measurement.
 If  a
system  has been prepared at  time $t_1$ in a state $|\Psi_1\rangle$ and  is
found
 at  time $t_2$
in a state $|\Psi_2\rangle$, then at time $t$, $t_1<t<t_2$, the system is
described
by
 $$
\langle \Psi_2 | e^{i\int_{t_2}^{t} H dt}
{\rm ~~~ and~~~}
e^{-i\int_{t_1}^{t} H dt} |\Psi_1\rangle . \eqno(7)
 $$
 For simplicity,  we shall consider the free Hamiltonian to
be zero; then, the system at time $t$ is described by the two states
 $
\langle \Psi_2 |
$ and
$
|\Psi_1\rangle
$. In order to obtain such a system, we prepare an ensemble of systems in
the state $ |\Psi_1\rangle$, perform a measurement of the desired variable
using
separate measuring devices for each system in the ensemble, and perform the
post-selection measurement. If the outcome of the post-selection was not
the desired result, we discard the system and the corresponding measuring
device. We look only at measuring devices corresponding to the systems
post-selected in the state $\langle \Psi_2 |$.

 The basic concepts of the  two-state   approach, the weak value of a physical
variable $A$ in the time
interval between pre-selection of the state $| \Psi_1 \rangle$ and
post-selection of the state $ | \Psi_2 \rangle$ is given by $$
 A_w \equiv {{\langle \Psi_2 | A | \Psi_1 \rangle}
\over {\langle \Psi_2 |\Psi_1 \rangle}} ~~~~.
\eqno(8)$$
  Let us present
the main idea  by way of  a simple example. We  consider, at time $t$, a
quantum system
that was prepared at time $t_1$ in the state $| B = b \rangle$ and that was
found at time $t_2$ in the state  $| C = c \rangle$, $t_1 < t <t_2$. The
measurements at times $ t_1$ and $t_2$ are complete measurements of, in
general, noncommuting variables $B$ and $C$.   The  free Hamiltonian is zero,
and therefore,  the first quantum state  at time $t$ is  $| B = b \rangle$.  In
the
two-state  approach we characterize the system at time $t$   by
backwards-evolving state
$\langle C = c| $ as well. Our  motivation for including the future state
is as follws: if  we know that  a
measurement of $C$ has been performed at time $t$ then the outcome is  $ C = c$
with probability 1. This intermediate measurement,
however, destroys our knowledge that $ B = b$, since the coupling of the
measuring device to the variable $C$ can change  $B$.
The idea of weak measurements is to make the coupling with the measuring device
sufficiently weak so  that $B$ does not change.  In fact, we require that both
quantum
states do
not  change, neither the usual
one
 $| B = b \rangle$ evolving towards the future nor $\langle C = c| $
evolving  backwards.

During  the whole time interval
between $t_1$ and $t_2$, both $B  = b$ and  $ C = c $
are  true (in some sense). But  then, $B + C = b + c$ must  also be true. The
latter
statement, however, might not have meaning in the standard quantum formalism
because
the sum of the eigenvalues $b + c$  might not be an eigenvalue of the operator
$ B + C$. An
attempt to measure $B + C$ using a  standard measuring procedure  will lead to
some change of the two quantum states and thus
the outcome will not be  $b + c$. A weak measurement, however, will
yield $b + c$.

 When the ``strong'' value of an observable is known with certainty, that is,
we know  the outcome  of an ideal (infinitely strong) measurement with
probability 1, the weak value is equal to the strong value. Let us analyze
the example above. The strong value of $B$ is $b$,  its eigenvalue.  The
strong value of $C$ is $c$, as  we know from
{\it retrodiction}.
{}From the definition (1) immediately follows:
$B_w = b$ and  $C_w = c$. However,   weak values, unlike  strong values, are
defined   not just for $B$ and $C$, but for {\it all} operators. The strong
value of the sum
$B + C$ when $[B, C] \neq 0$ is not defined, but the weak value of the sum
is defined: $(B + C)_w = b + c$.

The system  at time $t$ in a pre- and post-selected ensemble is defined by
two states, the usual one evolving from the time of the preparation
and the state  evolving backwards in time from the  post-selection.
We may neglect  the free Hamiltonian if the time between
the pre-selection and  the post-selection is very short.
Consider a system that has been pre-selected in a state $|\Psi_1 \rangle$ and
shortly afterwards post-selected in a state $|\Psi_2
\rangle$.  The weak value of any physical variable $A$ in the time
interval between the pre-selection and the post-selection is given by Eq. (8).
Let us show briefly how weak values emerge from a measuring procedure
with a sufficiently weak interaction.

  We consider a sequence of
measurements: a pre-selection of $|\Psi_{1} \rangle$, a (weak)
measurement interaction of the form of Eq. (1), and a
post-selection measurement finding the state $|\Psi_2 \rangle$.  The
state of the measuring device (which was initially in a Gaussian state) after
this sequence is given (up to
normalization) by
$$
\Phi (Q) = \langle \Psi_2 \vert
e^{-iPA}
\vert \Psi_1 \rangle e^{ -{{Q ^2} /{2\Delta ^2}}} ~~~~.
\eqno(9)
$$
After simple algebraic manipulation we can rewrite it (in the
$P$-representation) as
$$
\tilde \Phi (P) =   \langle \Psi_2
\vert \Psi_1 \rangle ~ e^{-i {A_w} P} ~
e^{-{{\Delta}^2 {P^2}} /{2}} ~ + ~\langle \Psi_2 \vert
 \Psi_1 \rangle  \sum_{n=2}^\infty {{(iP)^n}\over{n!}}
[(A^n)_w - (A_w)^n]   e^{ -{{\Delta ^2 P^2}} /{2}}~.\eqno(10)
$$
\noindent
If $\Delta$ is sufficiently large, then we
can neglect the second term of (10) when we Fourier transform
 back to the  $Q$-representation.  Large $\Delta$
corresponds to weak measurement in the sense that the
interaction Hamiltonian
(1) is small.  Thus, in the limit of weak measurement, the final state
of the measuring device (in the $Q$-representation) is
$$\Phi  (Q) = (\Delta^2 \pi )^{-1/4} e^{ -{{(Q - A_w)^2} /{2\Delta
^2}}}~~~~. \eqno(11)$$
This state represents a measuring device pointing to the weak value, $A_w$.

Although we have showed this result for a specific von Neumann model of
measurements, the result is completely general: any  coupling
of a pre- and post-selected system to a
 variable $A$, provided the coupling
is sufficiently weak, results in effective
coupling to $A_w$. This  weak coupling between  a  single system and the
measuring device
will not, in most cases, lead to a distinguishable shift of the pointer
variable, but collecting the results of  measurements on an  ensemble of
 pre- and post-selected systems will yield the weak values of
a measured variable to any desired precision.

When the strength of the coupling to the measuring device goes to zero, the
outcomes of the measurement invariably yield the weak value. To be more
precise, a measurement  yields the real part of the weak value. Indeed,
the weak value is,
in general,  a complex number, but its imaginary part will contribute only a
phase to the wave function of the measuring device in the position
representation of the pointer. Therefore, the imaginary part will not affect
the probability distribution
of the pointer position, which is what we see in a  usual measurement.
However, the imaginary part of the weak value also has physical meaning. It
expresses itself as a change in the conjugate momentum of  the pointer
variable.$^7$

Let us consider a measurement of a spin
component of a spin-1/2 particle. We shall consider a particle prepared in the
initial state
spin ``up'' in the $\hat{x}$ direction and post-selected to be ``up'' in the
$\hat{y}$ direction. At the intermediate time we  measure, weakly, the spin
component in the  $\hat{\xi}$ direction which is bisector of  $\hat{x}$ and
$\hat{y}$, that is,
$
\sigma_\xi =   (\sigma_x + \sigma_y)/\sqrt 2
$. Thus
${|}\Psi_1 \rangle =|{\uparrow_x} \rangle$,
$|\Psi_2 \rangle =|{\uparrow_y} \rangle$, and  the weak
value of $\sigma_\xi$  in this case is:
$$
(\sigma_\xi)_w =
 {{\langle{\uparrow_y} |\sigma_\xi  |{\uparrow_x} \rangle}\over
{\langle{\uparrow_y} |{\uparrow_x} \rangle}} =
 {1\over\sqrt 2}{{\langle{\uparrow_y} | (\sigma_x + \sigma_y)  |{\uparrow_x}
\rangle}
\over {\langle{\uparrow_y} |{\uparrow_x} \rangle}} = \sqrt 2
 ~~.
\eqno(12)$$
This value is, of course, ``forbidden'' in the standard interpretation
where a spin component can obtain the (eigen)values $\pm1$ only.

The Hamiltonian for measuring $\sigma_\xi$ is
$$
H = g(t) P \sigma_\xi~~~~.
\eqno(13)
$$
After the measuring interaction,   the quantum state of the system and the
pointer of the
 measuring device is
 $$
 \cos {(\pi/8)} |{\uparrow_\xi} \rangle e^{ -{{(Q-1)^2} /{2\Delta ^2}}} + \sin
{(\pi/8)} |{\downarrow_\xi} \rangle  e^{ -{{(Q+1)^2} /{2\Delta ^2}}}~~.
 \eqno(14)
$$
The probability  distribution of the pointer position, if it is observed now
without post-selection,  is the sum of the distributions
for each spin value. It is, up to normalization,
 $$
 prob(Q) = \cos^2{(\pi/8)} e^{ -{{(Q-1)^2} /{\Delta ^2}}} + \sin^2
{(\pi/8)}  e^{ -{{(Q+1)^2} /{\Delta ^2}}}~~.
 \eqno(15)
$$
In  the usual  strong measurement,  $\Delta \ll 1$. In this case, the
probability distribution  of the pointer is localized around $-1$ and $+1$ and
it is
strongly correlated to the values of the spin, $\sigma_z = \pm1$.

Weak measurement corresponds to a  $\Delta$ that is much larger than the range
of the eigenvalues, that is,  $\Delta \gg 1$. The pointer
distribution has a  large uncertainty, and it is peaked between the
eigenvalues, more precisely, at the expectation value $\langle{\uparrow_x}
|\sigma_\xi  |{\uparrow_x} \rangle = 1/\sqrt 2$.
An outcome of an individual measurement usually will not be close to
this number, but it can be found from an ensemble of such measurements. Note,
that we have not yet
considered the post-selection.

In  order to simplify the
analysis of  measurements on  the pre- and post-selected ensemble, let us
assume that we first make the post-selection
of the spin of the particle and only then look at the pointer of the device
that weakly measures $\sigma_\xi$. We must get  the same result as if we
first look at the outcome of the weak measurement,  make the
post-selection, and discard all readings of the weak measurement corresponding
to  the cases in which the result is  not $\sigma_y
=1$. The post-selected state of the particle in the $\sigma_\xi$ representation
is
$
|{\uparrow_y} \rangle =  \cos {(\pi/8)} |{\uparrow_\xi} \rangle -  \sin
{(\pi/8)}|{\downarrow_\xi} \rangle
$.
 The state of the measuring device after the post-selection of the spin
state is obtained by projection of (14) onto the post-selected state:
 $$
 \Phi (Q) ={\cal N} \Bigl(\cos^2 {(\pi/8)} e^{ -{{(Q-1)^2} /{2\Delta ^2}}} -
\sin^2
{(\pi/8)} e^{ -{{(Q+1)^2} /{2\Delta ^2}}}\Bigr)~~, \eqno(16)
$$
where ${\cal N}$ is a normalization factor. The probability distribution of the
pointer variable is given by
 $$
 prob(Q) ={\cal N}^2 \Bigl (\cos^2 {(\pi/8)} e^{ -{{(Q-1)^2} /{2\Delta ^2}}} -
\sin^2
{(\pi/8)} e^{ -{{(Q+1)^2} /{2\Delta ^2}}}\Bigr)^2~~ .
 \eqno(17)
$$
If the measuring interaction is strong, that is, $\Delta \ll 1$, then the
distribution is localized around the eigenvalues $\pm 1$ (mostly around 1
because the pre- and post-selected probability to find $\sigma_\xi =1$ is
more than 85\%), see Figs. $1a,1b$. But when the strength of the coupling is
weakened, that is, $\Delta$ is increased, the  distribution gradually changes
to a
 single broad peak around $\sqrt 2$, the weak value, see Figs. $1c-1e$.

The width of the peak is large and therefore each individual reading of the
pointer usually will be pretty far from $\sqrt 2$. The physical meaning of
the weak value, in this case, can be associated only with  an ensemble of pre-
and post-selected particles. The accuracy of defining the center of the
distribution goes as $1/\sqrt N$; thus, bu increasing $N$, the number of
particles in the ensemble, we can find the weak value with any desired
precision, see  Fig. $1f$.

\vskip .7 cm

\centerline{\bf PROTECTION OF A TWO-STATE VECTOR}

\vskip .2 cm

We are familiar with weak measurements performed on a single system. In fact,
the
first work on weak measurements$^3$ considered such a case. We have shown how a
single measurement of the spin component of a spin-$N$ system could yield
the ``forbidden'' value $\sqrt{2} N$ with the uncertainty $ \sqrt{N}$. This is
the weak value  of $S_\xi$ for the two-state vector $\langle S_y{=}N | |
S_x{=}N \rangle$. Another example that we have investigated is the
measurement of the
 kinetic energy of a tunneling particle.$^9$ We have shown  for any precision
of the measurement that we can ensure a negative value reading of the measuring
device by an appropriate choice of the post-selection state.

However, in these examples there is no measurement of two-state vector. If our
measuring device for the spin measurement shows $\sqrt 2 N$, we cannot deduce
that
our two-state vector is $\langle S_y{=}N |  |
S_x{=}N \rangle$. Indeed, there are many other  two-state vectors that yield
the same weak value for the spin component, but we cannot  even claim that we
have one of these vectors, because the probability to obtain the ``forbidden''
outcome $S_\xi=\sqrt 2 N$ due to a statistical error of the measuring
device is much higher. The same applies to the  measurement of  kinetic energy
of a tunneling particle. The negative value shown by  the measuring device
 usually is  due to a statistical error, and only in very
rare cases does it correspond to a  particle  ``caught'' in the
tunneling process.

We could try to use several  weak measurements on a single pre- and
post-selected system in order  to specify the
two-state vector. But in that case these measurements will change the
two-state vector. Therefore, as  in   the case of
the measurement of the forward evolving single-state vector of a single
system, we need a protection procedure.

At  first look,  it seems that protection of a two-state vector is
impossible. Indeed, if we add a potential that makes one state to be a
nondegenerate
eigenstate, then the other state, if it is different, cannot be an
eigenstate too (the states of the two-state vector cannot be orthogonal). The
Zeno-type protection does not work either: if we test
that the system is in one state then we know that it is not in another state.
But, nevertheless, protection of the two-state vector is possible, as we
will show next.

 The procedure for protection of a two-state vector of a given system is
carried out by coupling the system to another  pre- and post-selected
system. The protection procedure takes advantage of the fact that
weak values  might  acquire complex values. Thus, the effective
Hamiltonian of the protection might not be Hermitian. Non-Hermitian
Hamiltonians act  in different ways on quantum states evolving forward and
backwards in time. This allows simultaneous protection of two different
states (evolving in opposite time directions).

Let us start with the description of the protection of a  two-state
vector of a spin-1/2 particle considered  previously, $\langle{\uparrow_y}
|  |{\uparrow_x} \rangle $. The protection procedure uses an external   pre-
and post-selected
 system $S$ of a large  spin $N$ that is coupled to our spin via the
interaction:
$$
H_{prot} = - {\bf S \cdot \sigma}. \eqno(18)
$$
The external system is pre-selected in the state $|S_x {=} N\rangle$ and
post-selected in the state $\langle S_y {=} N|$, that is, it is described by
the two-state
vector $\langle S_y {=} N
|  |S_x {=} N \rangle $. When $N$ is large, and the
interaction with our spin-1/2 particle is not too strong, the latter cannot
change significantly the two-state vector of the protective system $S$, and
the spin-1/2 particle ``feels'' the effective Hamiltonian in which {\bf $S$} is
replaced by its weak value,
$$
{\bf S}_w = {{\langle S_y = N
|(S_x, S_y, S_z)  |S_x = N  \rangle} \over{\langle S_y = N
 |S_x = N  \rangle}} = (N, N, iN)  .\eqno(19)
$$
Thus, the effective protective Hamiltonian is:
$$
H_{eff} =- N( \sigma_x +  \sigma_y + i\sigma_z). \eqno(20)
$$
Straightforward calculations show that this (non-Hermitian) Hamiltonian has
two (non- \break
orthogonal) eigenstates:
$|{\uparrow_x} \rangle $ (with eigenvalue $-N$) and $|{\downarrow_y} \rangle $
(with eigenvalue $N$). This result provides a certain test of our approach.
When we
consider the original problem given by  the Hamiltonian (18), we can  easily
see that if we start in the
state $|{\uparrow_x} \rangle $  then all
following measurements of $\sigma_x$  must yield value 1, whereas if we
start with the state $|{\downarrow_y} \rangle $ then all
following measurements of $\sigma_y$ must yield the value $-1$.

However, for backward evolving states the effective Hamiltonian is the
hermitian conjugate of (20) and it has different eigenstates:  $\langle
{\uparrow_y}
|$ (with eigenvalue $-N$) and $\langle {\downarrow_x} |$ (with eigenvalue
$N$). Again, it is easily seen  that if  the particle is
post-selected in the state $\langle {\uparrow_y}
|$ then all preceding measurements of $\sigma_y$ must yield  $\sigma_y=1$,
whereas  the post-selection of $|{\downarrow_x} \rangle $ ensure $\sigma_x=-1$
for all preceding measurements.

The two-state vectors $\langle{\uparrow_y}
|  |{\uparrow_x} \rangle $ and $\langle{\downarrow_x}
|  |{\downarrow_y} \rangle $ do not change under the action of  the Hamiltonian
(18). In
order to prove that this Hamiltonian  indeed provides the protection, we
have to show that  measuring interactions with  the  spin components of the
particle will not lead to significant changes. For example, we must show that
we can
measure the weak value of $
\sigma_\xi =   (\sigma_x + \sigma_y)/\sqrt 2
$, which is $({\sigma_\xi})_w = \sqrt 2$, on a single particle. (As
previously shown, without
protection, the weak value is obtained only with an uncertainty that is
larger than the observed value;  therefore  in order to find the weak value
the pre- and post-selected  ensemble has to be used, see Fig. 1.) The effective
Hamiltonian during the measuring process is the sum of (1) and (20):
$$
H_{eff} =- N( \sigma_x +  \sigma_y + i\sigma_z) + {P\over \sqrt 2} ( \sigma_x +
 \sigma_y). \eqno(21)
$$
For any realistic measurement, $P$ is effectively bounded; thuus, for $N$ large
enough,
the second term will not change significantly the eigenvectors. The
two-state vector $\langle{\uparrow_y}
|   |{\uparrow_x} \rangle $ will remain essentially unchanged during the
measurement, and therefore the measuring device on this single particle
will yield $({\sigma_\xi})_w = \sqrt 2$. This weak value by itself is not
enough
to establish the two-state vector, but we can perform several weak
measurements such as $({\sigma_x})_w = 1$, $({\sigma_y})_w = 1$, and
$({\sigma_z})_w = i$ that uniquely
define the two-state vector.

We have shown that the Hamiltonian (18), with an  external system described
by the two-state vector $\langle S_y = N
|  |S_x = N \rangle $, provides protection for the two-state vector
$\langle{\uparrow_y}|  |{\uparrow_x} \rangle $. It is not difficult to
demonstrate
that any two-state vector obtained by pre- and post-selection of  the
spin-1/2 particle can be protected by the Hamiltonian (18). A general form of
the two-state vector is $\langle{\uparrow_\beta}|  |{\uparrow_\alpha}
\rangle $ where  $\hat{\alpha}$ and $\hat{\beta}$ denote some directions. It
can be
verified by a straightforward calculation that the two-state vector
$\langle{\uparrow_\beta}|  |{\uparrow_\alpha}
\rangle $ is protected when  the two-state vector of the protective device
is  $\langle S_\beta = N
|  |S_\alpha = N \rangle $.

One can naively suggest the following simple explanation of the above
procedure.  We pre-select the external system  in a state $|S_\alpha {=} N
\rangle$. Large $N$ corresponds to
the classical limit, so this is equivalent to a  ``magnetic'' field in the
$-\hat{\alpha}$
direction. Thus,  the quantum states of the system under study evolving to the
future ``feel'' this
strong magnetic field. The  state of the system,  $|{\uparrow_\alpha}
\rangle $, is a ground state and therefore it is protected. Similarly, for
the states evolving backwards in time, there is strong ``magnetic'' field in
$-\hat{\beta}$ direction, protecting the state $\langle{\uparrow_\beta}|$.
However, this picture is too naive. Based on this argument, one would
expect that in addition to  $|{\uparrow_\alpha}
\rangle $ the forward evolving state  $|{\downarrow_\alpha}
\rangle $ is  also protected, however, this is not so. There exists another
forward evolving protected
state, but it is $|{\downarrow_\beta}
\rangle $. Also, in addition to  $\langle {\uparrow_\beta}
| $ there exists another protected backward evolving state, however,  it
is $\langle{\downarrow_\alpha}|$ and not the expected state
$\langle{\downarrow_\beta}|$.

The failure of this naive
explanation does not allow a simple protection scheme of the two-state vector
of an
arbitrary quantum system, $\langle \Psi _2|  | \Psi _1 \rangle $. According
to this scheme, we construct a coupling of the system under study to the
external system  such that $| \Psi _1 \rangle $
is a ground state when the external system is in a state $| \Phi _1 \rangle $
and  $\langle \Psi _2 | $
is a ground (backward evolving) state when the external system is in a state
$\langle \Phi _2 | $. The  difficulty here, namely,  that the post-selection of
the state
$\langle \Psi _2 | $ is impossible because usually in this situation
$\langle \Phi _2 | \Phi _1 \rangle = 0 $,
cannot be naively solved by adding a tiny component of the pre-selected state
to the post-selected one, that is, post-selecting $\langle \Phi _2 | + \epsilon
\langle \Phi _1 |$ instead of $\langle \Phi _2 |$. Even for $\epsilon$  very
small, the backward evolving
state is not protected, and therefore the two-state vector is  not
protected either.

The proper way for protecting  a
two-state vector of an arbitrary system is  a generalization of  the protection
procedure of the two-state vector of a spin-1/2 particle described above.
 The task is to
protect a two-state vector
 $\langle \Psi _2|  | \Psi _1 \rangle $.
 Let us decompose the post-selected state
 $| \Psi _2 \rangle = a | \Psi
_1 \rangle + b | \Psi _\bot \rangle $.
 Now we can define ``model spin'' states:
$ | \Psi _1 \rangle \equiv | \tilde{\uparrow}_z \rangle$ and $ | \Psi _\bot
\rangle
\equiv | \tilde{\downarrow}_z \rangle$.
 On the basis of the two orthogonal states we
can obtain all other ``model spin'' states. For example,
$ |\tilde{\uparrow}_x \rangle = 1/\sqrt 2 ~( |\tilde{\uparrow}_z \rangle +
|\tilde{\downarrow}_z \rangle)$,
and then we can define  the ``spin model'' operator $\bf \tilde{\sigma}$. Now,
the
protection Hamiltonian, in complete analogy with the spin-1/2 particle case
is
$$
H_{prot} = - {\bf S \cdot \tilde{\sigma}}. \eqno(22)
$$
In order to protect the state $\langle \Psi _2|  | \Psi _1 \rangle $, the
pre-selected state of the external system has to be $|S_z {=} N \rangle$ and
the
post-selected state has to be  $\langle S_\chi {=} N|$ where the direction
$\hat{\chi}$ is defined by the  ``spin model'' representation of the state
$| \Psi _2\rangle$:
$$|\tilde{\uparrow}_\chi \rangle \equiv | \Psi _2\rangle = \langle \Psi _1
| \Psi _2 \rangle|\tilde{\uparrow}_z \rangle + \langle \Psi _\bot
| \Psi _2 \rangle|\tilde{\downarrow}_z \rangle . \eqno(23)
$$

For general quantum states  $ | \Psi _1 \rangle $ and $\langle \Psi _2|$,
the required protection  is a gedanken experiment. In general, the protection
Hamiltonian (22) generates
nonlocal interactions which can contradict relativistic causality. However,
what we investigate here is a conceptual question in the framework of
non-relativistic quantum theory, where any Hamiltonian is allowed.

\vskip .7 cm

\centerline{\bf CONCLUSIONS}

\vskip .2 cm

We have shown in the framework of nonrelativistic quantum theory that we
can measure (or, maybe a better word, ``observe'') two-state vectors
describing pre- and post-selected
quantum systems. A number of (non-ideal)
measurements define the two-state vector and we have a procedure to
protect the two-state vector from significant change due to these measurements.
In order to protect, we have to know the two-state vector. Thus, this
procedure is also liable to the criticism$^{11-12}$ leveled at  our first
proposal. Our response  to this can be found in Ref. 13.  Although we consider
our present proposal as a measurement
performed on a single system, it should also be
mentioned  that in any realistic practical implementation
 we will need  ensembles of particles, protective systems, and  measuring
devices.  The external system of the protective
device has to be not only prepared (pre-selected) in a certain state, but
also post-selected in a given state. In all interesting cases  the
probability for an appropriate outcome of the post-selection measurement is
extremely small. Still, there is a non-zero probability that our first run
with a single system, a single protective device, and a single set of measuring
devices will yield the desired outcomes. In this case we have a reliable
measurement
performed on a single system. However, even when  we  use  a pre-selected
ensemble,  we actually use only  a single pre- and post-selected
system. After achieving the first successful post-selection, we have
completed  the
experiment. For more discussion of this point, see Ref. 14.

It is interesting to notice that our procedure cannot protect a {\it
generalized two-state vector}$^8$ which is a superposition of two-state
vectors. The system described by a generalized two-state vector is correlated
to
some external system. It seems that it is impossible to find any protective
procedure of the generalized two-state vector that does not involve
coupling to that external system. This feature hints that the generalized
vector, although  useful as a tool, is not a basic concept. The composite
system consisting of the system under study and the system correlated to it  is
described by
the usual, basic two-state vector.

Let us come back to the questions raised in the Introduction: is there a
contradiction between ``reality'' of the Schr\"odinger wave, that is, the
single-state vector, and
``reality'' of the two-state vector? Our answer is that the complete
reality is described by the two-state vector. The single-state vector
gives a partial description when we  have only partial information.
The  apparent paradox of the  descriptions is as follows. Consider a spin-1/2
particle described by the two-state vector $\langle{\uparrow_y}
|  |{\uparrow_x} \rangle $. The value $\sigma_y$ corresponding to this
particle is $\sigma_y =1$. However, because  it is described by the single
(pre-selected)
forward evolving state $ |{\uparrow_x} \rangle $, the value of $\sigma_y
$ is considered as the expectation value, $\langle{\uparrow_x}
|\sigma_y  |{\uparrow_x} \rangle = 0 $. According to our claims both are
observable; thus,  how can they  be different?

In order to observe a quantum state it has to be protected. When we
discussed the protective experiments of single-state vectors we did not say
anything about quantum states evolving backwards in time. (It was not
related to the point we wanted to make.) However, the protective procedure
that we
proposed, automatically protects {\it identical} backward evolving state.
Thus, what we have proposed as an observation of a Schr\"odinger wave is indeed
an
observation a two-state vector with identical forward and backward evolving
states. For example, the protection of spin-1/2 particle state,$^2$ a strong
magnetic field in a given direction, protects the  two-state vector
with either  both states parallel or anti-parallel to this direction. This
procedure is
incompatible with the protection of the forward evolving state parallel to one
direction and the backward evolving state parallel to another. If the particle
is described by   $\langle{\uparrow_y}
|  |{\uparrow_x} \rangle $ then the strong magnetic field in
the $\hat{x}$ direction will change  of the backward
evolving spin-state. There exists a protection procedure for $ |{\uparrow_x}
\rangle $
that does not change the backward evolving state as  was described in the
preceding section. The ``observation'' of the state protected in such a
way will not yield the pre-selected quantum state but it will yield the
picture defined by the two-state vector.

Thus, the contradiction is resolved by giving a more accurate interpretation
of our original  protective measurement of the Schr\"odinger wave. We observed
not a
single-state vector, but a two-state vector with identical backward and
forward evolving states.

It is a pleasure to thank
 Sandu Popescu and Jacob Grunhaus  for helpful discussions. This research was
supported in part by  grant 425/92-1 of the
Basic
Research Foundation (administered by the Israel Academy of Sciences and
Humanities).

\vskip 0.7 cm

\centerline{\bf
 REFERENCES}
\vskip .2  cm

\noindent
1. Y. Aharonov and L. Vaidman, {\it Phys. Lett.} {\bf A178}, 38
(1993).\hfill \break
2. Y. Aharonov, J. Anandan, and L. Vaidman, {\it Phys. Rev.} {\bf
 A 47}, 4616 (1993).\hfill \break
3. Y. Aharonov, D. Albert, A. Casher, and L. Vaidman,
 {\it Phys. Lett.} {\bf  A 124}, 199 (1987).\hfil \break
4. L. Vaidman, Y. Aharonov and D. Albert
{\it Phys. Rev. Lett.} {\bf  58}, 1385 (1987).\hfill \break
5. Y. Aharonov,  D. Albert, and L. Vaidman,
{\it Phys. Rev. Lett.} {\bf  60}, 1351 (1988).\hfil \break
6. Y. Aharonov, J. Anandan, S. Popescu, and L. Vaidman,
{\it Phys. Rev. Lett.} {\bf  64}, 2965 (1990).\hfil \break
7.  Y.Aharonov and L. Vaidman, {\it Phys.  Rev.}   {\bf A 41}, 11 (1990).
\hfil \break
8.  Y.Aharonov and L. Vaidman, {\it J. Phys.} {\bf A 24}, 2315
(1991).\hfil \break
9. Y. Aharonov, P.G. Bergmann  and J.L.
   Lebowitz,   {\it  Phys.  Rev.}  {\bf B134},  1410 (1964).\hfil
\break
10. Y. Aharonov,  S. Popescu, D.~Rohrlich, and L.~Vaidman, {\it Phys.  Rev.}
{\bf A 48}, 4084 (1993).\hfil
\break
11.~~ W. G. Unruh, {\it Phys. Rev.} {\bf A 50}, 882 (1994). \hfill  \break
12.~~ C. Rovelli, {\it Phys. Rev.} {\bf A 50}, 2788 (1994). \hfill  \break
13. Y. Aharonov, J. Anandan, and L. Vaidman, The Meaning of the Protective
Measurements, TAUP-2195-94, E-board hep-th/9408153 (1994).\hfill \break
14. L. Vaidman,
in {\it Advances in Quantum Phenomena}, E. Beltrametti and J.M.
Levy-Leblond eds.,  NATO ASI series, Plenum, NY (1994).

\vfill
\break
\centerline{FIGURE CAPTIONS}
\vskip 1.5 cm

\noindent {\bf
Figure 1.~ Measurement on pre- and post-selected ensemble.}~  Probability
distribution of the pointer variable for measurement of
$\sigma_\xi$ when the particle is pre-selected in the state $\vert
{\uparrow_x} \rangle$ and post-selected in  the state $\vert
{\uparrow_y} \rangle$. The  strength of the
measurement  is
parameterized by the width of the distribution $\Delta$.
 ($a$) $\Delta = 0.1$; ($b$) $\Delta = 0.25$; ($c$) $\Delta =
1$; ($d$) $\Delta = 3$; ($e$) $\Delta = 10$.   ($f$) Weak measurement on the
ensemble
of 5000 particles; the original width of the peak, $\Delta = 10$, is reduced to
$10/\sqrt 5000 \simeq 0.14$. In the strong measurements ($a$)-($b$) the pointer
is localized
around the eigenvalues $\pm1$, while in the weak measurements ($d$)-($f$)
the peak of the distribution is located in the weak value $(\sigma_\xi)_w =
\langle {\uparrow_y}|
\sigma_\xi |{\uparrow_x} \rangle/\langle {\uparrow_y}|{\uparrow_x} \rangle
= \sqrt2$. The outcomes of the weak measurement on the ensemble
of 5000 pre- and post-selected particles, ($f$), are clearly outside the
range of the eigenvalues, $(-1,1)$.

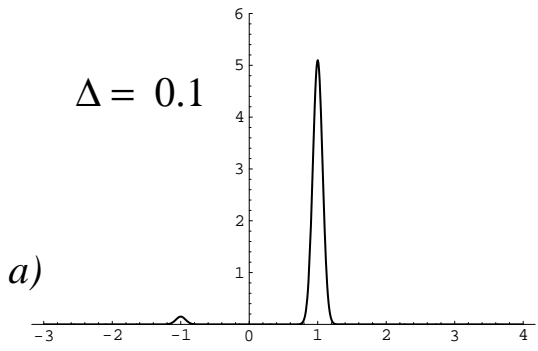
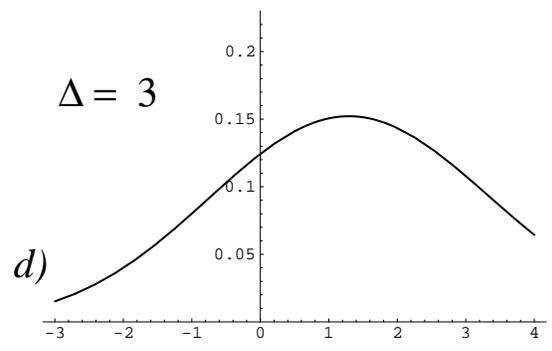
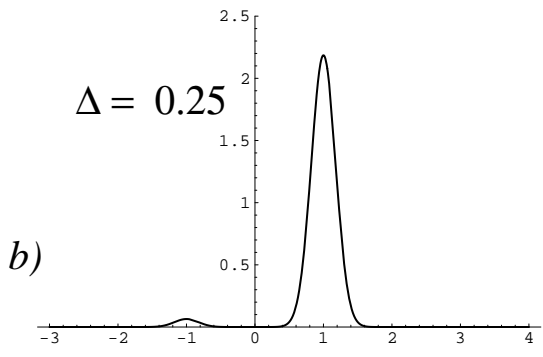
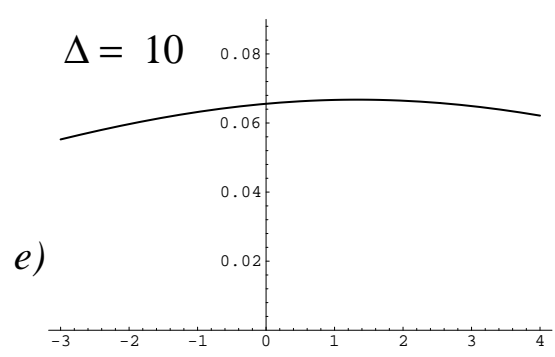
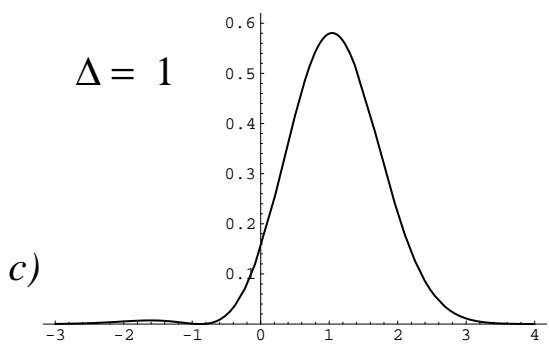
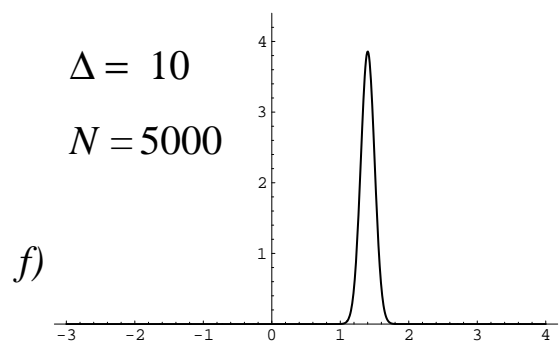
\end